# NONLINEAR OPTICS AS A PATH TO HIGH-INTENSITY CIRCULAR MACHINES*

S. Nagaitsev#, A. Valishev FNAL, Batavia, IL 60510, U.S.A
V. Danilov SNS, Oak Ridge, TN 37830, U.S.A.

*Abstract*

What prevents us from building super-high intensity accelerators? The answer is case-specific, but it often points to one of the following phenomena: machine resonances, various tune shifts (and spreads), and instabilities. These three phenomena are interdependent in all present machines. In this paper we propose a path toward alleviating these phenomena by making accelerators nonlinear. This idea is not new: Orlov (1963) and McMillan (1967) have proposed initial ideas on nonlinear focusing systems for accelerators. However, practical implementations of such ideas proved elusive, until recently [1].

## INTRODUCTION

All present accelerators (and storage rings) are built to have "linear" focusing optics (also called lattice). The lattice design incorporates dipole magnets to bend particle trajectory and quadrupoles to keep particles stable around the reference orbit. These are "linear" elements because the transverse force is proportional to the particle displacement, $x$ and $y$. This linearity results (after the action-phase variable transformation) in a Hamiltonian of the following type:

$$H(J_1, J_2) = \nu_x J_1 + \nu_y J_2, \quad (1)$$

where $\nu_x$ and $\nu_y$ are betatron tunes and $J_1$ and $J_2$ are actions. This is an integrable Hamiltonian. The drawback of this Hamiltonian is that the betatron tunes are constant for all particles regardless of their action values. It has been known since early 1960-s that the spread of betatron tunes is extremely beneficial for beam stability due to the so-called Landau damping. However, because the Hamiltonian (1) is linear, any attempt to add non-linear elements (sextupoles, octupoles) to the accelerator generally results in a reduction of its dynamic aperture, resonant behavior and particle loss. A breakthrough in understanding of stability of Hamiltonian systems, close to integrable, was made by N. Nekhoroshev [2]. He considered a perturbed Hamiltonian system:

$$H = h(J_1, J_2) + \varepsilon\, q(J_1, J_2, \theta_1, \theta_2), \quad (2)$$

where $h$ and $q$ are analytic functions and $\varepsilon$ is a small perturbation parameter. He proved that under certain conditions on the function $h$, the perturbed system (2) remains stable for an exponentially long time. Functions $h$ satisfying such conditions are called *steep* functions with quasi-convex and convex being the steepest. In general, the determination of steepness is quite complex. One example of a non-steep function is a linear Hamiltonian Eq. (1).

___________________________________________
*Work supported by UT-Battelle, LLC and by FRA, LLC for the U. S. DOE under contracts No. DE-AC05-00OR22725 and DE-AC02-07CH11359 respectively.
#nsergei@fnal.gov

In Ref. [1] we proposed three examples of nonlinear accelerator lattices. In this paper we will concentrate on one of the lattices, which we know results in a steep (convex) Hamiltonian. We will also describe how to implement such a lattice in practice.

## NON-LINEAR LATTICE

Consider an element of lattice periodicity consisting of two parts: (1) a drift space, $L$, with exactly equal horizontal and vertical beta-functions, followed by (2) an optics insert, $T$, which has the transfer matrix of a thin axially symmetric lens (Figure 1).

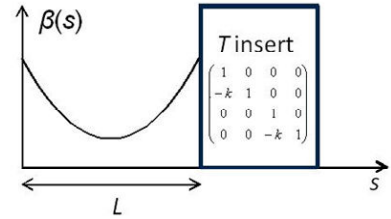

Figure 1: An element of periodicity: a drift space with equal beta-functions followed by a $T$ insert.

Let us assume that we have equal linear focusing in the horizontal and vertical planes such that the beta-functions in the drift space are equal to

$$\beta(s) = \frac{L - sk(L-s)}{\sqrt{1 - \left(1 - \frac{Lk}{2}\right)^2}}. \quad (2)$$

The insert $T$ can be implemented with regular elements (quadrupoles, dipoles, drifts) as described below. Let us now introduce additional transverse magnetic field along the drift space $L$. The potential, $V(x, y, s)$, associated with this field satisfies the Laplace equation, $\Delta V = 0$.

Now we will make a normalized-variable substitution [1] to obtain the following Hamiltonian for a particle moving in the drift space $L$ with an additional potential $V$:

$$H_N = \frac{p_{xN}^2 + p_{yN}^2}{2} + \frac{x_N^2 + y_N^2}{2} + U(x_N, y_N, \psi), \quad (3)$$

where

$$U(x_N, y_N, \psi) = \beta(\psi)V\!\left(x_N\sqrt{\beta(\psi)}, y_N\sqrt{\beta(\psi)}, s(\psi)\right) \quad (4)$$

and $\psi$ is the "new time" variable defined as the betatron phase,

$$\psi' = \frac{1}{\beta(s)}. \quad (5)$$

The potential $U$ in equation (4) can be chosen such that it is time-independent [1]. This results in a time-independent Hamiltonian (3). We will now choose a potential such that the Hamiltonian (3) possesses the second integral of motion. We will omit the subscript $N$ from now on.

Consider potentials [3] that can be presented in elliptic coordinates in the following way

$$U(x,y) = \frac{f(\xi) + g(\eta)}{\xi^2 - \eta^2}, \quad (6)$$

where $f$ and $g$ are arbitrary functions,

$$\xi = \frac{\sqrt{(x+c)^2 + y^2} + \sqrt{(x-c)^2 + y^2}}{2c}$$
$$\eta = \frac{\sqrt{(x+c)^2 + y^2} - \sqrt{(x-c)^2 + y^2}}{2c} \quad (7)$$

are elliptic variables and $c$ is an arbitrary constant.

The second integral of motion yields

$$I(x,y,p_x,p_y) = (xp_y - yp_x)^2 + c^2 p_x^2 + 2c^2 \frac{f(\xi)\eta^2 + g(\eta)\xi^2}{\xi^2 - \eta^2} \quad (8)$$

First, we would notice that the harmonic oscillator potential $(x^2 + y^2)$ can be presented in the form of Eq. (6) with $f_1(\xi) = c^2\xi^2(\xi^2 - 1)$ and $g_1(\eta) = c^2\eta^2(1 - \eta^2)$. Second, we have found the following family of potentials that satisfy the Laplace equation and, at the same time, can be presented in the form of Eq. (6):

$$f_2(\xi) = \xi\sqrt{\xi^2 - 1}(d + t\operatorname{acosh}(\xi)),$$
$$g_2(\eta) = \eta\sqrt{1 - \eta^2}(q + t\operatorname{acos}(\eta)) \quad (9)$$

where $d$, $q$, and $t$ are arbitrary constants. Thus, the total potential energy in Hamiltonian (3) is given by

$$U(x,y) = \frac{x^2}{2} + \frac{y^2}{2} + \frac{f_2(\xi) + g_2(\eta)}{\xi^2 - \eta^2}. \quad (10)$$

Of a particular interest is the potential with $d = 0$ and $q = \frac{\pi}{2}t$, because its lowest multipole expansion term is a quadrupole. Figure 2 presents a contour plot of the potential energy (15) for $c = 1$ and $t = 0.4$.

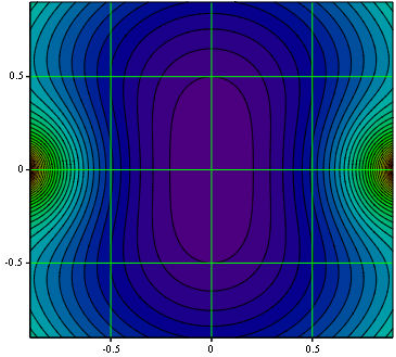

Figure 2: A contour plot of the potential energy Eq. (10) with $c = 1$ and $t = 0.4$.

The multipole expansion of this potential for $c = 1$ is as follows:

$$U(x,y) \approx \frac{x^2}{2} + \frac{y^2}{2} \quad (11)$$
$$+ t\operatorname{Re}\left((x+iy)^2 + \frac{2}{3}(x+iy)^4 + \frac{8}{15}(x+iy)^6 + \frac{16}{35}(x+iy)^8 + ...\right)$$

Since the 2D Hamiltonian with this potential has two integrals of motion, it is integrable and thus can be expressed as a function of actions:

$$H = h(J_1, J_2), \quad (12)$$

where

$$J_1 = \frac{1}{2\pi}\oint p_\eta d\eta \quad J_2 = \frac{1}{2\pi}\oint p_\xi d\xi \quad (13)$$

Let us now determine the maximum attainable betatron frequency spread in such a potential. First, this potential provides additional focusing in $x$ for $t > 0$ and defocusing in $y$. Thus, for a small-amplitude motion to be stable, one needs $0 \le t < 0.5$. This corresponds to the following small-amplitude betatron frequencies,

$$\nu_1 = \nu_0\sqrt{1+2t}$$
$$\nu_2 = \nu_0\sqrt{1-2t}, \quad (12)$$

where $\nu_0$ is the unperturbed linear-motion betatron frequency. For arbitrary amplitudes the frequencies are obtained by

$$\nu_1(J_1, J_2) = \frac{\partial h}{\partial J_1}$$
$$\nu_2(J_1, J_2) = \frac{\partial h}{\partial J_2}. \quad (13)$$

Figure 3 presents frequencies $\nu_1(J_1, 0)$ and $\nu_2(0, J_2)$, normalized by $\nu_0$ for $t = 0.4$.

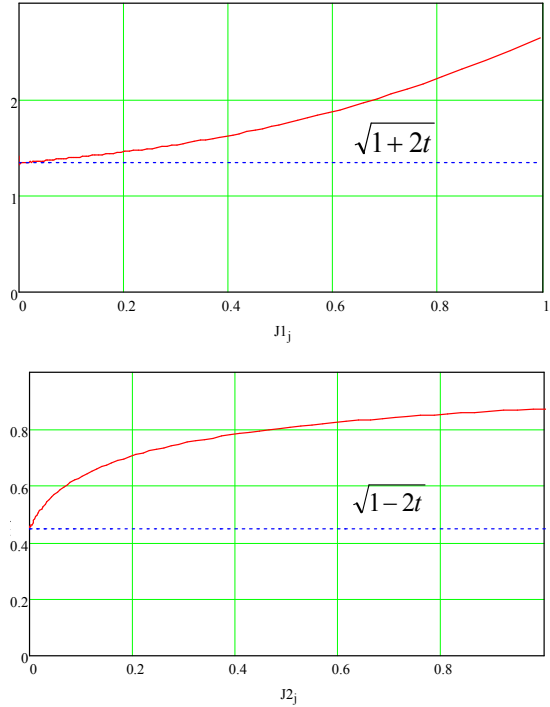

Figure 3: Oscillation frequencies $\nu_1(J_1, 0)$ (top) and $\nu_2(0, J_2)$ (bottom), normalized by $\nu_0$, for $t = 0.4$

By examining the function $h$ in Eq. (12) one can also demonstrate that it is a convex function and thus satisfies the Nekhoroshev's condition for a steep Hamiltonian. In the next section we discuss how to implement such a system in a practical accelerator.

## PRACTICAL IMPLEMENTATION

Since only a part of the accelerator circumference must be occupied by the nonlinear elements, it is natural to start with a conventional design machine. The lattice must satisfy the following design criteria:

- Be periodic, with the element of periodicity comprised of a drift space with equal beta-functions, and a focusing and bending block with the betatron phase advance in both planes equal to $\pi$ (T-insert in Fig. 1).
- The T-insert must be tunable to allow a wide range of phase advances (and beta-functions) in the drift space in order to study different betatron tune working points.
- It is preferable that the focusing block is achromatic in order to avoid strong coupling between the transverse and longitudinal degrees of freedom.

Currently, a superconducting RF test facility is under construction at Fermilab's New Muon Lab [4]. Upon completion, the facility will consist of an electron linac delivering bunches with the energy of up to 750MeV and an experimental area located in a 16x16 m hall. The experimental program for NML includes advanced accelerator physics R&D, and a small storage ring for studies of nonlinear dynamics could be included as a part of that program. Considering the NML hall space and beam energy constraints, we restricted the machine to approx. 13x13 m footprint (Fig. 4).

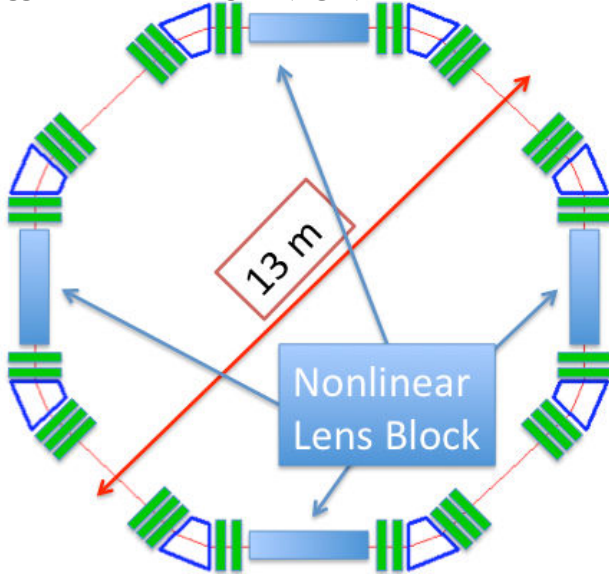

Figure 4: Layout of the test ring.

In the design of the test ring, the lattice has four periods, in which a Double Bend Achromat with 10 quadrupoles represents the T-insert. The drifts for the nonlinear lens blocks have a length of 3 m. There are also four 2.5 m straight sections for installation of an RF cavity, injection devices and instrumentation. The lattice functions of the periodicity element are presented in Fig. 5, and main parameters of the machine are listed in Table 1.

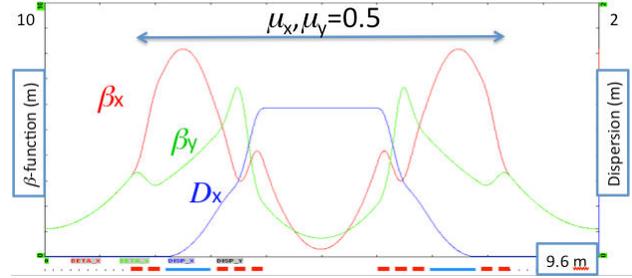

Figure 5: Test ring lattice functions.

| $e^-$ Energy | 150 MeV |
|---|---|
| Circumference | 38 m |
| Dipole field | 0.5 T |
| Betatron tunes $Q_x=Q_y$ | 2.4÷3.6 |
| Synchrotron radiation damping time | 1-2 s ($10^7$ turns) |
| Transverse emittance (rms non-normalized) | $6\times10^{-8}$ m |

Table 1: Main parameters of the test ring.

Such machine can be used to test the nonlinear integrable optics concept by demonstrating stable operation at super-high values of the betatron tune spread. In the proposed lattice design, the phase advance $\nu_0$ over the drift space with nonlinear element can be varied from 0.1 to 0.4 (this corresponds to the betatron tune between $(0.5+0.1)\times4=2.4$ and $(0.5+0.4)\times4=3.6$). According to (12), the maximum attainable tune spread in this case can exceed 1, which means that some particles within the bunch would cross the integer resonance.

In order to demonstrate the high tune spread, the transverse beam size must be comparable to the distance between the poles of the potential $U$ (Fig. 2). For the chosen ring energy and equilibrium emittance, the beam size $\sigma_x,\sigma_y\approx0.25$ mm, which would require an impractically small transverse dimensions of the nonlinear elements. However, due to the very long damping time it is possible to "paint" a larger area with the small emittance linac beam. Hence, we considered nonlinear elements with the aperture $2c\geq2$ cm.

It is not practical to realize the continuous variation of the cross section of the nonlinear element as required by (3). Rather, one would construct the nonlinear lens block of a number of elements with constant cross section. This modification presents a perturbation of the ideal integrable system. In addition, the integrability can be disturbed by optics errors common to conventional accelerators, such as the beta-function and phase advance modulation. These factors motivated the study of the system stability using numerical simulation.

Macro particle tracking codes were used to simulate the effect of various factors on the stability of particle motion.

The simulations also generate the dipole moment spectra, a quantity that can be used to evaluate the betatron tune spread and which is reported by common accelerator instrumentation.

In the simulation, the nonlinear lenses were implemented as thin kicks, and tracking through the accelerator arcs was performed with conventional methods. A typical simulation would track 5000 particles over 8,000 turns to produce the spectra and $10^6$ turns to check the particle stability. The initial distribution had the amplitude of particles limited by $c/2$ in the horizontal plane and $c$ in the vertical plane, and random phases

Figure 6 presents the dipole moment spectra for the case of the ring betatron tunes $Q_x$=3.6, $Q_y$=3.62 ($\nu_0$=0.4) and different magnitude of nonlinearity $t$. As one would expect at $t$=0 there is no tune spread since the machine lattice is linear. The tune spread grows as the nonlinearity increases. For $t$=0.4 the maximum tune spread is $\nu_0 \times 4 \times 1$=1.6 (see Fig. 3), which can not be seen in Fig. 6 due to the properties of the Fourier transformation, and because the observed quantity, the horizontal or vertical dipole moment, is a combination of normal modes. Some particles of the bunch had their tune on the integer resonance and yet no instability was observed even though the lattice was not perfectly symmetrical.

A more convenient presentation is shown in Fig. 7, where the spectra of horizontal dipole moment are plotted for a special initial particle distribution with $y$, $p_y$=0. For such case, the horizontal coordinate coincides with one of the normal modes and it is possible to compare the tracking results with the analytical model in Fig. 3. Indeed, for $t$=0.1 the tune for small amplitude particles is $0.5\times 4+\nu_0\sqrt{(1+2t)}$=3.75, and particles with larger amplitudes have a positive tune shift. For $t$=0.4, the small amplitude tune is $0.5\times 4+\nu_0\sqrt{(1+2t)}$=4.15, which at the plot is seen as 1-0.15=0.85.

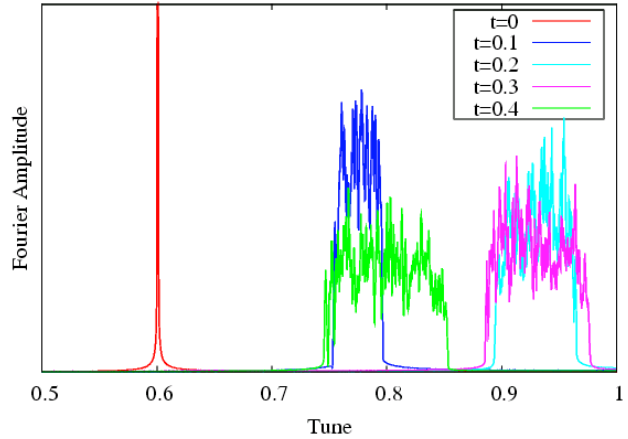

Figure 7: Spectrum of horizontal dipole moment for various values of nonlinearity $t$. $Q_y$=3.62, $y$, $p_y$=0.

The stability of the system to the following perturbations was studied:
- Phase advance in the T-insert not equal to $\pi$, different horizontal and vertical phase advance, differences between the elements of periodicity. It was found that up to 0.05 tune difference is tolerable.
- Different $\beta$-functions in the nonlinear lens blocks. Up to 5% variation between $x$ and $y$ did not cause particle losses.
- Misalignment of thin nonlinear elements within the lens block. Up to 5 cm error in the longitudinal position of the individual element is allowed, although the tolerance depends on the phase advance in the nonlinear straight section and on the value of nonlinearity. The system is less sensitive to perturbations at smaller values of $\nu_0$ and $t$.

A more elaborate study of the system stability range is underway with the focus on machine nonlinearities, such as the chromaticity correction sextupoles, and the effect of longitudinal dynamics, e.g. the importance of chromaticity of T-inserts and zero dispersion in the nonlinear lens section.

## SUMMARY

In this paper we presented an example of completely integrable non-linear optics and its practical implementation.

Tune spreads of 50% are possible. In our test ring simulation we achieved tune spread of about 1.5 (out of 3.6).  Such a system has the potential to make an order of magnitude increase in beam brightness and intensity because of increased Landau damping.

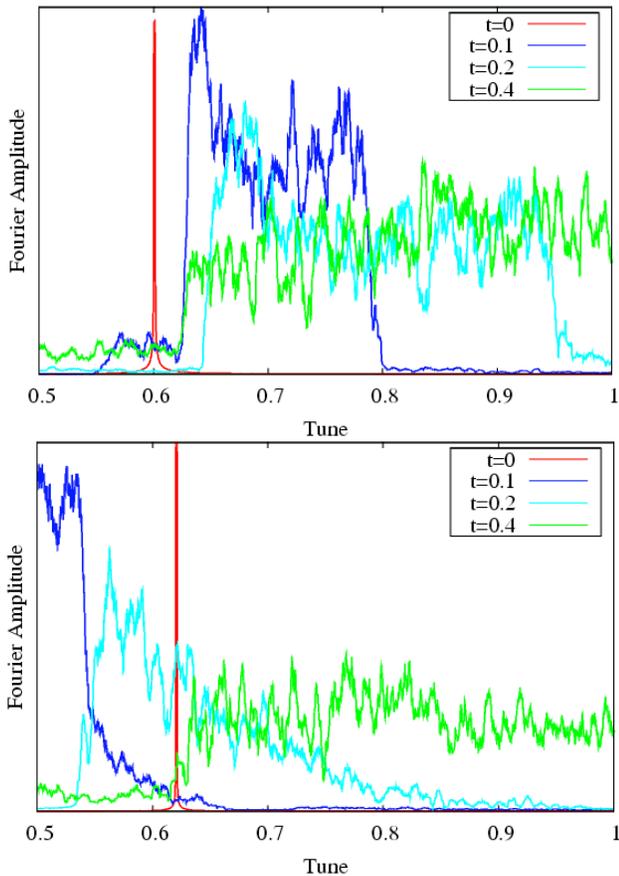

Figure 6: Spectra of horizontal (upper plot) and vertical (lower) dipole moment for various values of nonlinearity $t$. Linear ring betatron tunes $Q_x$=3.6, $Q_y$=3.62.